\documentclass[useAMS,usenatbib]{mn2e}

\newcommand{\dev}{\ensuremath{\,\mathrm{d}}}
\usepackage{amssymb}
\usepackage{amsmath}
\usepackage[dvips]{graphicx}
\usepackage{epsfig}
\usepackage{multicol}

\title[Gamma-Ray Burst afterglow scaling coefficients for general density profile]{Gamma-Ray Burst afterglow scaling coefficients for general density profiles}
\author[H.J. van Eerten and R.A.M.J. Wijers]{H.J. van Eerten$^{1}$\thanks{E-mail:
H.J.vanEerten@uva.nl (HJvE); R.A.M.J.Wijers@uva.nl (RAMJW)} and R.A.M.J. Wijers$^{1}$\footnotemark[1]\\
$^{1}$Astronomical Institute 'Anton Pannekoek', Kruislaan 403, 1098 SJ Amsterdam, the Netherlands}
\begin{document}

\date{Accepted . Received ; in original form }

\pagerange{\pageref{firstpage}--\pageref{lastpage}} \pubyear{200x}

\maketitle

\label{firstpage}

\begin{abstract}
Gamma-ray burst (GRB) afterglows are well described by synchrotron emission originating from the interaction between a relativistic blast wave and the external medium surrounding the GRB progenitor. We introduce a code to reconstruct spectra and light curves from arbitrary fluid configurations, making it especially suited to study the effects of fluid flows beyond those that can be described using analytical approximations. As a check and first application of our code we use it to fit the scaling coefficients of theoretical models of afterglow spectra. We extend earlier results of other authors to general circumburst density profiles. We rederive the physical parameters of GRB 970508 and compare with other authors.
\end{abstract}

\begin{keywords}
gamma rays: bursts -- gamma rays: theory -- plasmas -- radiation mechanisms: nonthermal -- shockwaves 
\end{keywords}

\section{Introduction}
In the fireball model, Gamma-Ray Burst (GRB) afterglows are thought to be the result of synchrotron radiation generated by electrons during the interaction of a strongly collimated relativistic jet from a compact source with its environment (for recent reviews, see \citealt{Piran2005, Meszaros2006}). Initially the resulting spectra and light curves have been modelled using only the shock front of a spherical explosion and a simple power law approximation for the synchrotron radiation (e.g. \citealt{Wijers1997, Meszaros1997, Sari1998, Rhoads1999}).  One or more spectral and temporal breaks were used to connect regimes with different power law slopes. For the dynamics the self similar approximation of a relativistic explosion was used \citep{Blandford1976}. These models have been refined continuously. More details of the shock structure were included (e.g. \citealt{Granot1999, Gruzinov1999}), more accurate formulae for the synchrotron radiation were used (e.g. \citealt{Wijers1999}) and efforts have been made to implement collimation using various analytical approximations to the jet structure and lateral spreading behaviour (see \citealt{Granot2005} for an overview). On top of that, there have been studies focussing on arrival time effects (e.g. \citealt{Huang2007}) and some numerical simulations (e.g. \citealt{Salmonson2003, Granot2001, Nakar2007}).

The aim of this paper is twofold. The first aim is to introduce a new method to derive light curves and spectra by post-processing relativistic hydrodynamic (RHD) jet simulations of arbitrary dimension, properly taking into account all beaming and arrival time effects, as well as the precise shape of the synchrotron spectrum and electron cooling (in this paper we  will ignore self-absorption, although it can in principle be included in our method). This is done in sections \ref{peak_section} and \ref{cooling_section}.

The second aim is to present a set of scaling coefficients for the slow-cooling case for a density profile $\rho = \rho_0 \cdot (R/R_0)^{-k}$ for general values of $k$. Fits to afterglow data using $k$ as a free fitting parameter have yielded values markedly different from both $k=0$ and $k=2$ \citep{Starling2007}, although with error bars not excluding either option. The scaling coefficients have been obtained by application of our post-process code not to a full hydrodynamic simulation but to an emulation of this. From the spherical Blandford \& McKee (BM) analytical solution for the blast wave for the impulsive energy injection scenario, snapshots containing the state of the fluid at given emission times were constructed and stored to provide the input for the post-process code.

The use of the BM solution provides us with an opportunity to check the results and the consistency of the code in an environment where we already have a lot of analytical control and understanding. The scaling coefficients are presented in section \ref{coefficients_section}. They can be used by observers to obtain the physical parameters for the blast wave (e.g. explosion energy and circumburst density) from the values for the peak flux and break frequencies that have been obtained from fits to the data. Readers interested only in the coefficients can skip ahead to this point. The fluxes in the transitional regions between the different power law regimes have often been described using heuristic equations that smoothly change from one dominant power law to the next. The abruptness of this change depends on a sharpness parameter $s$. Using the detailed results from our simulations, in section \ref{smoothPL_section} we provide equations for $s$ in terms of two fit parameters: the slope of the accelerated particle distribution $p$ and the aforementioned $k$ that describes the circumburst density structure. In section \ref{application_section} we apply our results to GRB 970508. We discuss our results in section \ref{discussion_section}. Some cumbersome equations and derivations have been deferred to appendices.

\section{Description of the post-processing code}
\label{peak_section}
The code takes as input a series of snapshots of relativistic hydrodynamics
configurations on a (in this paper, one-dimensional) grid. Although we will treat only the analytical Blandford-McKee solution \citep{Blandford1976} for the blast wave dynamics put on a grid here, the code is written with the intention to interact with the AMRVAC adaptive mesh refinement code \citep{Meliani2007} and will read from file the following conserved variables:
\begin{equation}
D = \gamma \rho', \quad \vec{S} = \gamma^2 h' \vec{v}, \quad \tau = \gamma^2 h' - p' - \gamma \rho' c^2,
\end{equation}
with $\gamma$ the Lorentz factor, $\rho'$ the proper density, $h'$ the relativistic (i.e. including rest mass) enthalpy density, $\vec{v}$ the proper velocity, $p'$ the pressure and $c$ the speed of light. From the conserved values we can reconstruct all hydrodynamical quantities using the equation of state
\begin{equation}
p' = ( \Gamma_{ad} - 1 ) e_{th}',
\end{equation}
where $\Gamma_{ad}$ the adiabatic index that is kept fixed and $e_{th}'$ the thermal energy density. In the entire paper, all comoving quantities will be primed.

The grids represent a spherically symmetric fluid configuration and all grid cells are assumed to emit a fraction of their energy as radiation. This fraction of course has to be small enough not to affect the dynamics, since the post-processing approach does not allow for feedback. For the time being we restrict ourselves to the optically thin case.

Four ignorance parameters are provided to the code at runtime: $p$, $\xi_N$, $\epsilon_E$ and $\epsilon_B$, denoting respectively the slope of the relativistic particle distribution, the fraction of particles accelerated to this relativistic distribution at any given time, the fraction of thermal energy that is carried by the relativistic electrons and the fraction of thermal energy that resides in the (tangled-up) magnetic field. To be precise: the fractions $\epsilon_E$ and $\epsilon_B$ are fractions of $e_{th}'$, which is strictly speaking the sum of the thermal energy of the protons and non-accelerated electrons plus the energy of the accelerated electrons plus the magnetic field energy. Since we consider fully relativistic gases, the adiabatic indices of the electrons and protons are both at $\Gamma_{ad} = 4/3$. Also, if the magnetic flux enclosed by the surface of any arbitrary fluid element is an adiabatic invariant, we find that $B^2 \propto \rho^{4/3}$, which tells us that the behaviour of the magnetic energy density $B^2/8\pi$ is identical to that of the thermal energy. Or in other words, $\epsilon_B$ retains a constant value away from the shock front. The fraction of shock-accelerated particles $\xi_N$ is often set to one, but we have already kept it explicit in our calculations. At late times (i.e. when the fluid flow is no longer relativistic) $\xi_N$ has te be lower than unity in order to have enough energy per accelerated particle for synchrotron emission.

In this work we consider synchrotron radiation only. All grid cells contain a macroscopic number of radiating particles and the radiation from these particle distributions is calculated following \citet{Sari1998} and \citet{ Rybicki}, but with two important differences: the transition to the lab frame is postponed as long as possible and no assumption about the dynamics of the system is used anywhere as this should be provided by the snapshot files. 

For clarity of presentation we will ignore the effect of electron cooling in this section.

For the emitted power per unit frequency of a typical electron we have
\begin{eqnarray}
 \frac{ \dev P'_{<e>}}{\dev \nu'} (\nu') & = & \frac{p-1}{2} \cdot \frac{\sqrt{3}{q_e}^3B'}{m_ec^2} \cdot 
    Q \left( \frac{\nu'}{\nu'_{cr,m}}\right).
\label{ensemble_electronpower_equation}
\end{eqnarray}
Here $q_e$ denotes the electron charge, $m_e$ the electron mass (later on we will also encounter the proton mass $m_p$) and $B'$ the local magnetic field strength. The function $Q$ contains the shape of the spectrum. It shows the expected limiting behaviour: $Q(x) \propto x^{1/3}$ for $x \ll 1$ and $Q(x) \propto x^{(1-p)/2}$ for $x \gg 1$.
It incorporates an integration over all pitch angles between electron velocities and the local magnetic field and an integration over the accelerated particle distribution. We use a power law particle distribution with a lower cut-off Lorentz factor $\gamma_m$. Equation (\ref{ensemble_electronpower_equation}), the critical frequency $\nu'_{cr,m}$ and the full shape of $Q$ are derived in appendix \ref{distro_section}.

Assuming isotropic radiation in the comoving frame, we arrive at
\begin{equation}
\frac{ \dev^2 P_{<e>}' } {\dev \nu' \dev \Omega' } ( \nu' ) = \frac{1}{4 \pi} \frac{ \dev P'_{<e>}  }{\dev \nu'}(\nu')
\end{equation}
per solid angle $\Omega'$.

To get to the \emph{received} power per unit volume in the lab frame, we have to apply the correct beaming factors, Doppler shift the frequency and multiply the above result for a single particle with the lab frame particle density:
\begin{equation}
 \frac{ \dev^2 P_V }{\dev \nu \dev \Omega}( \nu' (\nu) ) = \frac{\xi_N n}{\gamma^3 (1-\beta \mu)^3} \cdot \frac{ \dev^2 P_{<e>}' } {\dev \nu' \dev \Omega' } ( \nu \gamma (1 - \beta \mu)  ),
\end{equation}
with $\mu$ now denoting the cosine of the angle between the fluid velocity and the observer (unprimed, so measured in the lab frame), $\beta$ the fluid velocity in units of $c$ and $n$ the number density.

Finally, the flux the observer receives at a given observer time is given by
\begin{equation}
 F(\nu) = \frac{1}{r_{obs}^2} \int \frac{ \dev^2 P_V }{\dev \nu \dev \Omega}( \nu' (\nu) ) ( 1 - \beta \mu ) c \dev A \dev t_e.
\label{integral}
\end{equation}
Here $r_{obs}$ is the observer distance\footnote{For cosmological distances $r_{obs}$ denotes the luminosity distance and redshift terms $(1+z)$ need to be inserted in the appropiate places in the equations.}, approximately the same for all fluid cells (though the differences in arrival times \emph{are} taken into account). The area $A$ denotes the \emph{equidistant surface}. For every emitting time $t_e$ a specific intersecting (with the radiating volume) surface exists from which radiation arrives exactly at $t_{obs}$. The integration over the emission times $t_e$ (represented in the different snapshot files) requires an extra beaming factor and a factor of $c$ to transform the total integral to a volume integral.

To perform the surface integrals, the post-processing code uses a Monte Carlo integration algorithm with both importance and stratified sampling, using the pseudo-random Sobol' sequence.\footnote{But if symmetry allows (e.g. the observer is on the jet axis), we just do a straightforward Bulirsch-Stoer integration} For the integral over emission times, a combination of modified midpoint integration and Richardson extrapolation is used (the latter allowing us to occasionally skip a snapshot if the desired convergence is already reached). All methods are explained in detail in \citet{Press1992}. A minor complication is here that not every $t_e$ probed has a corresponding snapshot file available and interpolation between snapshot files may be needed. The boundaries for surface $A$ are analytically known conic sections and depend on the jet opening angle and observer angle. Two useful consistency checks are observing a spherical explosion from different angles and calculating the volume of a grid snapshot via integration over different observer times while setting the emissivity to one.

When creating snapshot files directly from the BM solution we found that sufficient convergence (below the cooling break) was obtained during the post-processing even for modest grid resolutions.\footnote{On the order of 120 base cells with 8 levels of refinement (an increase in refinement means a local increase of resolution by a factor of two) for a region $\backsim 10^{17}$ cm to $\backsim 10^{18}$ cm and a relatively small number of snapshots ($\backsim 1000$) to go from $\Gamma \backsim 100$ down to $\Gamma \backsim 2$. Unfortunately, the resolution will eventually be dictated by that required by RHD simulations, which will be much higher.}

For spherical explosions we used jets with an opening angle of 180 degrees, which makes no noticeable difference for the resulting signal because of relativistic beaming. It is worth emphasizing that it is our method that allows for the modest grid resolution and keeps calculation time short. This is because instead of binning the output from all grid cells, it takes an observer time as the starting point and then probes the appropriate contributing grid cells only (resolving the structure within the cell by including neighbouring cells in the interpolation). We have checked our results by increasing the accuracy (e.g. larger number of grid cells, more snapshot files, smaller step sizes in the integrals etc.) and by replacing the Monte Carlo integration routine with a nested one-dimensional Bulirsch-Stoer algorithm. These consistency checks are in addition to the two mentioned earlier. Finally we have checked the grid interpolation and snapshot I/O routines by comparing the results of our post-process code with those of a code that does not read profiles from disc but calculates the BM solution at run time.

\section{The inclusion of electron-cooling}
\label{cooling_section}
The code as described so far is purely a post-process code that in principle can be applied directly to the output of any RHD simulation. If we want to include electron cooling however, we can no longer reconstruct the electron energy distribution from the conserved quantities alone. In the particle distribution function, in addition to the lower boundary $\gamma_m$ , we will also have an upper boundary $\gamma_M$ beyond which all electrons have cooled.
The time evolutions of both the lower cut-off Lorentz factor $\gamma'_m$ and the upper cut-off Lorentz factor $\gamma'_M$ (that we have tacitly kept at infinity in the previous section) of this distribution are no longer dictated by adiabatic cooling alone but also by radiation losses. This implies that when running an RHD simulation we need to keep track of at least one extra quantity (at least $\gamma'_M$, although in practice we will trace both).

With the introduction of a second critical frequency $\nu'_{cr,M}$, the equation describing the total emitted power now becomes, 
\begin{equation}
 \frac{ \dev P'_{<e>}}{\dev \nu'} (\nu') = \frac{p-1}{2} \cdot \frac{\sqrt{3}{q_e}^3B'}{m_ec^2} \cdot 
    \mathcal{Q} \left( \frac{\nu'}{\nu'_{cr,M}}, \frac{\nu'}{\nu'_{cr,m}}\right),
\end{equation}
instead of eq. (\ref{ensemble_electronpower_equation}). The function $\mathcal{Q}(x_M, x_m)$ and $\nu'_{cr,M}$ are derived and described in appendix \ref{cooling_details_section}. For $\gamma_M$ at infinity we have $\mathcal{Q}( 0, x_m ) \to Q(x_m)$.

The particle distribution that lies beneath the derivation of this new function $\mathcal{Q}$ is no longer a simple power law, but drops off sharply for particle Lorentz factors approaching the peak value of $\gamma'_M$. A subtlety worth noting here is that the critical frequency $\nu'_{cr,M}$ corresponding to $\gamma'_M$ is \emph{not} the cooling frequency, but a frequency beyond which the signal will drop exponentially. Since we put $\gamma'_M$ at infinity directly behind the shock, we will not directly observe $\nu'_{cr,M}$. The actual cooling frequency is found between $\nu'_{cr,m}$ and $\nu'_{cr,M}$, at the point where the shape of the particle distribution ceases to be characterized by a power law but starts to be characterized by the strong drop towards $\gamma'_M$. We will discuss the distinction between the cooled and uncooled region in appendix \ref{hot_region_section}.

A consequence of electron cooling is that the amount of energy in the shock-accelerated electrons is no longer a constant fraction of the thermal energy. $\epsilon_E$ now refers to the fraction of thermal energy in the shock accelerated electrons \emph{directly} behind the shock front instead and the further evolution of the available energy is traced via $\gamma_m$ and $\gamma_M$.

\section{Scaling coefficients}
\label{coefficients_section}
\begin{figure}
\begin{center}
\includegraphics[angle=0, width=0.2\textwidth]{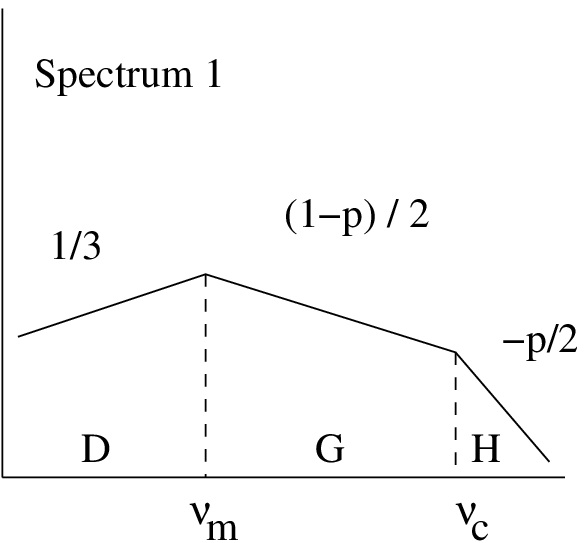}
\hspace{10mm}
\includegraphics[angle=0, width=0.2\textwidth]{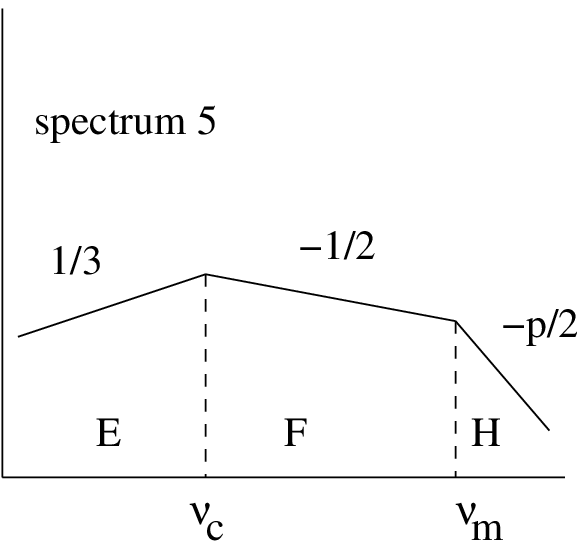}
\caption{different possible spectra}
\label{twospectra}
\end{center}
\end{figure}
Especially for high Lorentz factors, the shape of the spectrum is dominated by the radiation coming from a very thin slab right behind the shock front. So we expect the flux to scale as
\begin{equation}
 F  \propto (p-1) \cdot N_{tot} \cdot (\frac{  \dev \mu}{(1-\beta \mu)^3 \gamma^3} \cdot B' \cdot \mathcal{Q} \left( \frac{\nu}{\nu_{cr,M}},  \frac{ \nu }{\nu_{cr,m}} \right).
\label{F_scaling_equation}
\end{equation}
Here $N_{tot}$ is the total number of radiating particles and $\dev \mu$ reflects the increasing visible size (due to decrease of beaming) of the slab. The two possible spectra that the code can generate are shown in fig. \ref{twospectra}, where we used the labelling from \citet{Granot2002} to distinguish the different power law regimes. In tables (\ref{scalings_table}) and (\ref{frequencies_table}) we give the expressions for the absolute scalings in the different regimes $D$, $E$, $F$, $G$, $H$ and the critical frequencies. Scaling coefficients aside, these equations are similar to those given in \citet{vanderHorst2008}. The flux in regime $D$ is denoted by $F_D$, the critical peak frequency in spectrum 1 is denoted by $\nu_{m,1}$, the critical cooling frequency in spectrum 1 by $\nu_{c,1}$ and so on.

\begin{table*}
 \centering
 \begin{minipage}{140mm}
  \caption{Flux Scalings for the different regimes (see tables \ref{coefficients_table} and \ref{coefficient_values_table} for $C_D, C_E$ etc.)}
\begin{eqnarray*}
\hline
\hline
  F_D & = & C_D(p,k) \cdot \frac{\xi_N}{r_{obs,28}^2} \cdot \left( \frac{\epsilon_E}{\xi_N} \right)^{-2/3} \cdot \epsilon_B^{1/3} \cdot n_0^{\frac{2}{4-k}} \cdot E_{52}^{\frac{10-4k}{3(4-k)}} \cdot t_{obs,d}^{\frac{2-k}{4-k}} \cdot (1+z)^{\frac{10-k}{3(4-k)}} \cdot \nu^{1/3} \textrm{ mJy,} \nonumber \\
\hline
 F_E & = & C_E(p,k) \cdot \frac{ \xi_N }{ r_{obs,28}^2 } \cdot \epsilon_B \cdot n_0^{\frac{10}{3(4-k)}} \cdot E_{52}^{\frac{-6k+14}{3(4-k)}}
 \cdot t_{obs,d}^{\frac{2-3k}{3(4-k)}} \cdot (1+z)^{\frac{14-k}{3(4-k)}} \cdot \nu^{1/3} \textrm{ mJy.} \\
\hline
 F_F & = & C_F(p,k) \cdot \frac{ \xi_N }{ r_{obs,28}^2 } \cdot \epsilon_B^{-1/4} E_{52}^{3/4} \cdot t_{obs,d}^{-1/4} \cdot (1+z)^{\frac{3}{4}} \cdot \nu^{-1/2} \textrm{ mJy.} \\
\hline
 F_G & = & C_G(p,k) \cdot \frac{\xi_N}{r_{obs,28}^2} \cdot \left( \frac{\epsilon_E}{\xi_N} \right)^{p-1} \cdot \epsilon_B^{(p+1)/4} \cdot n_0^{2/(4-k)} \cdot E_{52}^{\frac{-kp-5k+4p+12}{4(4-k)}} \cdot t_{obs,d}^{\frac{3kp-5k-12p+12}{4(4-k)}} \\ \\
 & & \cdot (1+z)^{\frac{12-k+4p-kp}{4(4-k)}} \cdot \nu^{-(p-1)/2} \textrm{ mJy.} \\
\hline
F_H & = & C_H(p,k) \cdot \frac{\xi_N}{ r_{obs,28}^2 } \cdot \left( \frac{\epsilon_E}{\xi_N} \right)^{p-1} \cdot \epsilon_B^{(p-2)/4} \cdot E_{52}^{(p+2)/4} \cdot t_{obs,d}^{(2-3p)/4} \cdot (1+z)^{\frac{2+p}{4}} \cdot \nu^{-p/2} \textrm{ mJy.} \\
\hline
\hline
\end{eqnarray*}
\label{scalings_table}
\end{minipage}
\end{table*}

\begin{table*}
  \centering
\begin{minipage}{140mm}
 \caption{Critical frequencies for the different regimes (see tables \ref{coefficients_table} and \ref{coefficient_values_table} for $C_D, C_E$ etc.)}
\begin{eqnarray*}
\hline
\hline
\nu_{m,1} & = & \left( \frac{C_G}{C_D} \right)^{6/(3p-1)} \cdot \left( \frac{\epsilon_E}{\xi_N} \right)^2 \cdot \epsilon_B^{1/2} \cdot E_{52}^{1/2} \cdot t_{obs,d}^{-3/2} \cdot (1+z)^{1/2} \textrm{ Hz.} \\
\hline
\nu_{c,1} & = & \left( \frac{ C_H }{C_G} \right)^2 \cdot\epsilon_B^{-3/2} \cdot n_0^{\frac{-4}{4-k}} \cdot E_{52}^{\frac{3k-4}{2(4-k)}} \cdot t_{obs,d}^{\frac{-4+3k}{2(4-k)}} \cdot (1+z)^{-\frac{4+k}{2(4-k)}} \textrm{ Hz.} \\
\hline
\nu_{c,5} & = & \left( \frac{C_F}{C_E} \right)^{6/5} \cdot \epsilon_B^{-3/2} \cdot n_0^{-4/(4-k)} \cdot E_{52}^{\frac{3k-4}{2(4-k)}} \cdot t_{obs,d}^{\frac{-4+3k}{2(4-k)}} \cdot (1+z)^{-\frac{4+k}{2(4-k)}} \textrm{ Hz.} \\
\hline
\nu_{m,5} & = & \left( \frac{C_H}{C_F} \right)^{2/(p-1)} \cdot \left( \frac{ \epsilon_E }{\xi_N} \right)^2 \cdot \epsilon_B^{1/2} E_{52}^{1/2} \cdot t_{obs,d}^{-3/2} \cdot (1+z)^{1/2} \textrm{ Hz.} \\
\hline
\hline
\end{eqnarray*}
\label{frequencies_table}
\end{minipage}
\end{table*}

The equations in the tables introduce a number of symbols that need an explanation. The cosmic redshift is given by $z$, while the luminosity distance $r_{obs,28}$ is measured in units of $10^{28}$ cm. $E_{52}$ is the explosion energy $E$ in units of $10^{52}$ erg. The observer time in days is denoted by $t_{obs,d}$. The characteristic distance $R_0$ we put at $10^{17}$ cm and $\rho_0$ and $n_0$ are related via the proton mass: $\rho_0 = m_p n_0$. The scaling coefficients $C_D$, $C_E$ etc. contain a number of numerical constants (determined by fitting to output from our code) and some explicit dependencies on $k$ and $p$ and are further explained in appendix \ref{scaling_derivation_section}.

Before the cooling break the scaling behaviour is dictated by the asymptotic behaviour of $Q(\nu'/\nu'_{cr,m})$. The steepening of the spectrum beyond the cooling breaks and the corresponding changes in the scaling behaviour are due to the fact that beyond the cooling break frequency the region behind the shock that still significantly contributes to the total flux (i.e. the \emph{hot region}) becomes noticably smaller than the shock width. The changes in the scalings reflects the change in the size of region. The hot region is discussed separately in appendix \ref{hot_region_section}.

\section{sharpness of broken power law}
\label{smoothPL_section}

In simple power law model fits, the gradual transition between regimes is often handled by a free parameter, the sharpness factor $s$. In more detailed calculations like those done here the gradual transitions are included automatically and we can use this to provide the correct dependence of $s$ on $p$ and $k$. This eliminates $s$  as a free parameter, simplifying the fit to the data and allowing the shape of the transition to help determine whether a particular model fits the data or not.

For spectrum 1, we use the following equation to describe the flux density near the peak break $\nu_{m,1}$:
\begin{equation}
F(\nu) = F_{m,1} \cdot \left[ \left( \frac{\nu}{\nu_{m,1}} \right)^{\textstyle -\frac{s_{m,1}}{3}} + \left( \frac{\nu}{\nu_{m,1}} \right)^{\textstyle -\frac{s_{m,1}(1-p)}{2}} \right]^{\textstyle -\frac{1}{s_{m,1}}},
\label{Fsmooth_m1_equation}
\end{equation}
where $F_{m,1}$ denotes the flux at the critical frequency $\nu_{m,1}$ for infinite sharpness $s_{m,1}$ (i.e. the meeting point of the asymptotic power laws). When we switch off cooling in our simulation, we can determine $s_{m,1}$ from fitting against the resulting spectrum while keeping the other parameters in equation (\ref{Fsmooth_m1_equation}) fixed. The sharpness is a function mainly of $p$ and to a lesser extent of $k$ and the other simulation input parameters. Rather than attempting to include all secondary dependencies when formulating a description for $s_{m,1}$, we find that the following approximation for $s_{m,1}$ is always valid up to a few percent:
\begin{equation}
 s_{m,1} = 2.2 - 0.52 p.
\end{equation}
When we switch on electron cooling, the flux is best approximated by
\begin{eqnarray}
F(\nu) & = & F_{m,1} \nonumber \\
 & & \cdot \left[ \left( \frac{\nu}{\nu_{m,1}} \right)^{\textstyle -\frac{s_{m,1}}{3}} + \left( \frac{\nu}{\nu_{m,1}} \right)^{\textstyle -\frac{s_{m,1}(1-p)}{2}} \right]^{\textstyle -\frac{1}{s_{m,1}}} \nonumber \\
 & & \cdot \left[ 1 + \left( \frac{\nu}{\nu_{c,1}} \right)^{s_{c,1}/2} \right]^{\textstyle -\frac{1}{s_{c,1}}}.
\label{F1smooth_cooling_equation}
\end{eqnarray}
If we fit this function against simulation output using $s_{c,1}$ as a fitting parameter we find that the results are described (up to a few percent) by
\begin{equation}
s_{c,1} = 1.6 - 0.38 p -0.16 k + 0.078 pk. 
\end{equation}
A simultaneous fit using both $s_{m,1}$ and $s_{c,1}$ yields the same results.

For spectrum 5 the order of the breaks is reversed and the smooth power law for both breaks is given by
\begin{eqnarray}
F(\nu) & = & F_{c,5} \cdot \left[ \left( \frac{\nu}{\nu_{c,5}} \right)^{\textstyle -\frac{s_{c,5}}{3}} + \left( \frac{\nu}{\nu_{c,5}} \right)^{\textstyle \frac{s_{c,5}}{2}} \right]^{\textstyle -\frac{1}{s_{c,5}}} \nonumber \\
 & & \cdot \left[ 1 + \left( \frac{\nu}{\nu_{m,5}} \right)^{\textstyle s_{m,5}\cdot \frac{p-1}{2}} \right]^{\textstyle -\frac{1}{s_{m,5}}},
\end{eqnarray}
where $F_{c,5}$ denotes the peak flux for infinite sharpness $s_{c,5}$ and the prescriptions for the sharpness are
\begin{equation}
 s_{c,5} = 0.66 - 0.16k,
\end{equation}
and
\begin{equation}
s_{m,5} = 3.7 - 0.94p + 3.64k - 1.16pk.
\end{equation}
Once again valid up to a few percent. Given their accuracies, all sharpness prescriptions are consistent with \cite{Granot2002}.

\section{application to GRB 970508}
\label{application_section}
Various authors have used flux scaling equations to derive the physical properties of GRB 970508 from afterglow data \citep{Galama1999, Granot2002, Yost2003, vanderHorst2008}. This provides us with a context to illustrate the scaling laws derived in section \ref{coefficients_section}. We will use the fit parameters obtained from broadband modeling by \citet{vanderHorst2008}. They have fit simultaneously in time and frequency while keeping $k$ as a fitting parameter. Because the only model dependencies that have been introduced by this approach are the scalings of $t$ and $\nu$ (and no scaling coefficients), their fit results are still fully consistent with our flux equations. Using the cosmology $\Omega_M = 0.27$, $\Omega_\Lambda=0.73$ and Hubble parameter $H_0=71$ km s$^{-1}$ Mpc$^{-1}$, they have $r_{obs,28} = 1.635$ and $z =0.835$ \citep{Metzger1997}, leading, at $t_{obs,d} = 23.3$ days, to $\nu_{c,1} = 9.21\cdot10^{13}$ Hz, $\nu_{m,1} = 4.26\cdot10^{10}$ Hz, $F_{m,1} = 0.756$ mJy, $p = 2.22$ and $k = 0.0307$. 

Both \citet{vanderHorst2008} and \citet{Galama1999} take for the hydrogen mass fraction of the circumburst medium $X=0.7$, which in our flux equations is mathematically equivalent (though conceptually different) to setting $\xi_N = (1+X)/2 = 0.85$. Unfortunately this still leaves us with four variables to determine ($\epsilon_B$, $\epsilon_E$, $E_{52}$, $n_0$) and only three constraints (peak flux, cooling and peak frequency). From a theoretical study of the microstructure of collisionless shocks \citet{Medvedev2006} arrives at the following constraint:
\begin{equation}
\epsilon_E \backsim \sqrt{\epsilon_B}.
\end{equation}
We include this constraint to have a closed set of equations.

\begin{figure}
\begin{center}
\includegraphics[angle=-90, width=0.49\textwidth]{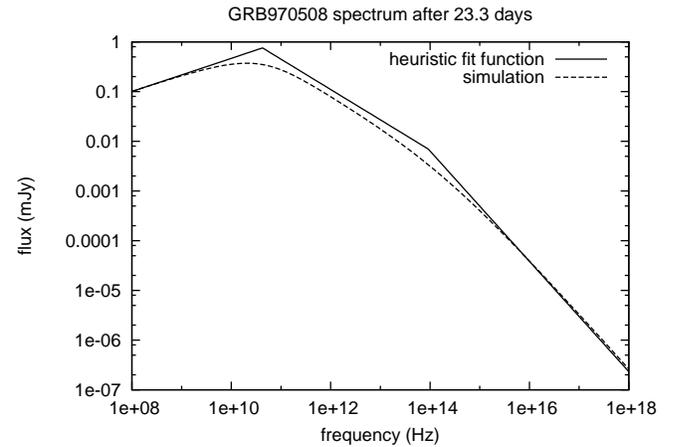}
\caption{A comparison of a representation of the data, using the values of Van Der Horst for the critical frequencies and peak flux, combined with our equations for the transition sharpnesses, at 23.3 days, with a reproduced spectrum from a BM blast wave simulation.} 
\label{GRB970508_spectrum_figure}
\end{center}
\end{figure}

\begin{figure}
\begin{center}
\includegraphics[angle=-90, width=0.49\textwidth]{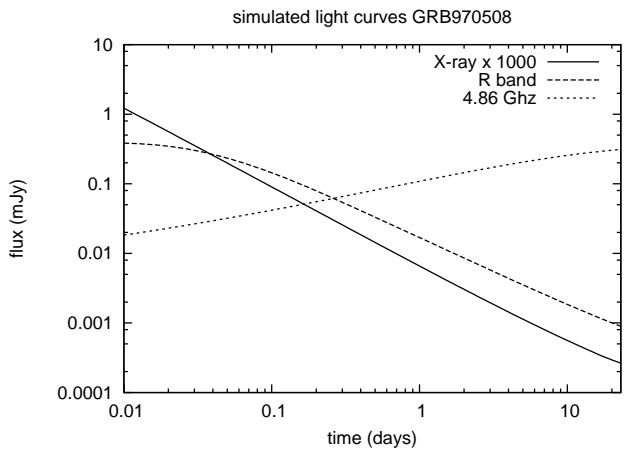}
\caption{Simulated light curves generated from post-processing of simulations using the BM solution with the input parameters we have derived: $E_{52} = 0.155$, $n_0 = 1.28$, $\epsilon_B = 0.105$, $\epsilon_E = 0.325$. For the X-rays we used $\nu = 1\cdot10^{18}$ Hz and for the R band $\nu = 4.3 \cdot 10^{14}$ Hz.}
\label{GRB970508_lightcurves_figure}
\end{center}
\end{figure}

For the values quoted above we obtain: $E_{52} = 0.155$, $n_0 = 1.28$, $\epsilon_B = 0.1057$, $\epsilon_E = 0.325$. In figures \ref{GRB970508_spectrum_figure} and \ref{GRB970508_lightcurves_figure} we have plotted a comparison between the spectrum generated by using these values as input parameters for the BM solution and the spectrum as it is represented by applying the results of the broadband fit of \citet{vanderHorst2008} for the values of the critical frequencies and the peak flux to equation (\ref{F1smooth_cooling_equation}).

Our scaling coefficients were fixed for arbitrary $k$ and for comparison we also give results for $k = 0$ and $k=2$. The ISM case is virtually identical to $k=0.0307$ and yields: $E_{52} = 0.155$, $n_0 = 1.23$, $\epsilon_B = 0.106$, $\epsilon_E = 0.325$. The stellar wind case yields: $E_{52} = 0.161$, $n_0 = 6.45$, $\epsilon_B = 0.0957$, $\epsilon_E = 0.309$. The quantity $n_0$ (the particle number density at the characteristic distance $1\cdot10^{17}$  cm) is affected most.

Also for comparison we give some of the values obtained by other authors. \citet{Galama1999} obtain for the ISM case: $E_{52} = 3.5$, $n_0 = 0.03$, $\epsilon_B = 0.09$, $\epsilon_E = 0.12$. \citet{Granot2002} obtain for the ISM case: $E_{52} = 0.12$, $n_0 = 22$, $\epsilon_B = 0.012$, $\epsilon_E = 0.57$. Both use $p=2.2$. Finally \citet{vanderHorst2008} obtain for $k=0.0307$: $E_{52} = 0.435$, $n_0 = 0.0057$, $\epsilon_B = 0.103$, $\epsilon_E = 0.105$. 

The large differences between the various results illustrate the importance of using the correct scaling coefficients to derive physical parameters of GRBs and provide a strong motivation for this work. Because the error on $\epsilon_B$ in particular is rather large for the quoted authors, who have used the self absorption critical frequency to provide a fourth constraint, the constraint from \citet{Medvedev2006} can not be rejected based on their fit results. The extension of our code to include self-absorption will yield an alternative and can be used to further study the applicability of Medvedev's constraint. 

\section{Summary and discussion}
\label{discussion_section}
In this paper we have introduced an approach to reconstruct light curves and spectra from hydrodynamic simulations. The central idea is that we do not start from simulation snapshots and bin the output of each grid cell, but that for representative snapshots we integrate over the intersecting surface that contains all points where radiation is generated that is due to arrive at a given observer time and frequency. When performing these integrations we interpolate within and between grid cells. While in the context of this paper we have used only snapshots that contain mimicked RHD output using the BM solution, first results using real simulations have been obtained and will be discussed in a later paper. An important thing to note here is that, even though the post-process code only required a very modest resolution, the underlying hydrodynamics code usually does not. \citet{Meliani2007-2} used 1200 base level cells and 15 refinement levels to simulate the evolution of the blast wave (earlier, when they were putting their code to the test they even used 30,000 base level cells at one point, see \citealt{Meliani2007}). This means that, in general, parallel computer systems are required to run these simulations, something for which the RHD code that our post-process code interacts with (AMRVAC) was explicitly designed.

In our code we included synchrotron radiation and electron cooling. We use a parametrisation of the accelerated particle distribution in terms of $\gamma_M$ and $\gamma_m$. Thermal radiation from the particles not accelerated to a power law distribution can be included in a straightforward manner. The code can also be extended to include self-absorption and since the outgoing synchrotron radiation from a grid cell is independent of the incoming radiation, this can be done without expanding to a full radiative transfer code including scattering. Effectively, all that is needed is to postpone the integration over the intersecting surfaces until after the integration over emission times, while in the meantime diminishing the output from earlier surfaces according to the column densities in the lines of sight, which amounts to solving linear transport equations only.

As a consistency check and a first application of the code we calculated the scaling coefficients of the flux scaling equations for GRB afterglow spectra for arbitrary values of $k$ with unprecedented accuracy. These results can be used to obtain the physical parameters of the burst from fits to afterglow data. For the ISM and stellar wind scenario's the results have been checked against the results of \citet{Granot2002} and are found to be fully consistent. The motivation for the choice of arbitrary $k$ is that various authors have now used $k$ as a fitting parameter (e.g. \citealt{vanderHorst2008, Yost2003}). Values of $k$ other than 0 or 2 reflect the structure of a circumburst medium altered by shock interactions or more complicated stellar wind structures. We have used GRB970508 to illustrate the effect of using our scaling coefficients to deduce the physical properties of a GRB. Here we have used an additional constraint by \citet{Medvedev2006} to obtain a closed set of equations in the absence of a full description for the self-absorption.

\section*{Acknowledgments}

This research was supported by NWO Vici grant 639.043.302 (RAMJW) and NOVA project 10.3.2.02 (HJvE). HJvE wishes to thank Atish Kamble and Alexander van der Horst for useful discussion. We are indebted to the anonymous referee for pointing out a numerical error in our original submission (the origin of this type of error is discussed in appendix \ref{hot_region_section}).

\appendix
\onecolumn
\section{Derivation of emitted power per electron}
\label{distro_section}
For each electron Lorentz factor $\gamma_e$ we define two critical frequencies $\nu_{cr,e, \alpha}'$ and $\nu_{cr,e}$: 
\begin{equation}
\nu_{cr,e, \alpha}' = \frac{3}{4\pi} \gamma_{e}'^2 \frac{q_e B'}{ m_e c } \sin \alpha \equiv \nu_{cr,e}' \sin \alpha,
\label{nume}
\end{equation}
where $q_e$ denotes the electron charge, $m_e$ the electron mass and $\alpha$ the pitch angle between field and velocity. It is around (but not exactly \emph{at}) these values that the spectrum peaks and we will find them useful as integration variables later on.

The power per unit frequency emitted by an electron is (\cite{Rybicki}):
\begin{equation}
\frac{ \dev P'_{e,\alpha} }{\dev \nu'} (\nu') = \frac{\sqrt{3} {q_e}^3 B' \sin \alpha}{m_e c^2} F(\frac{\nu'}{\nu'_{cr,e,\alpha}}),
\end{equation}
where
\begin{equation}
 F(x) \equiv x \int_x^\infty K_{\frac{5}{3}}(\xi) \dev \xi,
\end{equation}
with $K_{\frac{5}{3}}$ a modified Bessel function of fractional order. $F(x)$ behaves as follows in the limits of small and large $x$:
\begin{equation}
 F(x) \backsim \frac{4\pi}{\sqrt{3}\Gamma(\frac{1}{3})}\left(\frac{x}{2} \right)^{1/3} \left(1 - \frac{\Gamma(\frac{1}{3})}{2^{5/3}} x^{2/3} + \frac{3}{16} x^2 \right), \qquad x \ll 1,
\end{equation}
\begin{equation}
 F(x) \backsim \sqrt{ \frac{\pi}{2} } x^{1/2} e^{-x} \left( 1 + \frac{55}{72} \frac{1}{x} - \frac{10151}{10368} \frac{1}{x^2} \right), \qquad x \gg 1,
\end{equation}
where $\Gamma(x)$ is the gamma function of argument $x$.

For the mean power averaged over all pitch angles while assuming an isotropic pitch angle distribution we obtain:
\begin{equation}
 \frac{ \dev P'_{e} }{\dev \nu'} (\nu') = \frac{\sqrt{3} {q_e}^3 B'}{m_e c^2} \mathcal{P} (\frac{\nu'}{\nu'_{cr,e}}),
\label{Pe}
\end{equation}
where
\begin{equation}
 \mathcal{P}(x) \equiv \frac{1}{2} \int_0^{\pi} (\sin \alpha)^2 F(\frac{x}{\sin \alpha}) \dev \alpha.
\end{equation}
In the limit of small and large $x$, $P(x)$ behaves as follows:
\begin{equation}
 \mathcal{P}(x) \backsim \frac{2^{2/3} \cdot \pi^{3/2} \cdot \sqrt{3}}{9 \Gamma(\frac{11}{6})} x^{1/3} - \frac{\pi}{\sqrt{3}}x - \frac{5 \cdot \pi \cdot \sqrt{3} }{ 48 \cdot 2^{1/3} \Gamma( \frac{11}{6} )} x^{7/3}, \qquad x \ll 1,
\end{equation}
\begin{equation}
 \mathcal{P}(x) \backsim \frac{\pi}{2} e^{-x}, \qquad x \gg 1.
\end{equation}

The effective lower cut-off Lorentz factor of a collection of electrons $\gamma_m'$ can be expressed in terms of local fluid quantities. The integrated power law particle distribution $C \gamma_e'^{-p} \dev \gamma_e'$ ($C$ is a constant of proportionality) must yield the total number density of particles:
\begin{equation}
\int_{\gamma_m'}^{\infty} C (\gamma_e')^{-p} \dev \gamma_e' = \xi_N n' \to C = - \frac{1-p}{(\gamma_m')^{1-p}} \xi_N n'.
\label{electron_power_equation}
\end{equation}
Similarly the integrated particle energies must yield the total energy:
\begin{equation}
  \frac{\int_{\gamma_m'}^\infty C \gamma_e'^{-p} \ \gamma_e' m_e c^2 \ d \gamma_e'}{\int_{\gamma_m'}^{\infty} C \gamma_e'^{-p} d\gamma_e'} = \frac{ \epsilon_E e_{th}' + \xi_N n' m_e c^2 }{\xi_N n'}.
\end{equation}
Combining these equations and dropping the rest mass term in the energy equation (it will be negligible for relativistic electrons), we obtain
\begin{equation}
  \gamma_m' = \left( \frac{2-p}{1-p} \right) \cdot \left( \frac{\epsilon_E}{\xi_N} \frac{e_{th}'}{n'} \frac{1}{m_e c^2} \right).
\label{gamma_m_equation}
\end{equation}

If we integrate (\ref{Pe}) over the particle distribution and divide the result by the total electron density, we obtain the emitted power per ensemble electron\footnote{an ensemble electron contribution is therefore constructed as the total of all electron contributions divided by the number of electrons.}:
\begin{equation}
 \frac{ \dev P'_{<e>}}{\dev \nu'} (\nu') = \frac{p-1}{2} \cdot \frac{\sqrt{3}{q_e}^3B'}{m_ec^2} \cdot 
    Q \left( \frac{\nu'}{\nu'_{cr,m}}\right).
\end{equation}
Here $\nu'_{cr,m}$ denotes the resulting value of $\nu'_{cr,e}$ when we substitute $\gamma'_m$ for $\gamma'_e$ in equation (\ref{nume}). It surfaces when we switch integration variables from $\gamma'_e$ to $\nu'_{cr,e}$. The auxiliary function $Q$ is defined as
\begin{equation}
 Q ( x ) \equiv x^{\frac{1-p}{2}} \int_0^x y^{\frac{p-3}{2}} \mathcal{P} (y) \dev y.
\end{equation}
In the limit of small and large $x$, $Q(x)$ behaves as follows:
\begin{equation}
 Q(x) \backsim \frac{2^{5/3} \sqrt{3\pi} \Gamma( \frac{1}{6} )}{5(3p-1)} x^{1/3} - \frac{2\pi}{\sqrt{3}(p+1)} x + \frac{3 \sqrt{3\pi} \Gamma( \frac{1}{6} )}{2^{1/3}(88+24p)} x^{7/3}, \qquad x \ll 1,
\label{Q_small_equation}
\end{equation}
\begin{equation}
 Q(x) \backsim \sqrt{\pi} \frac{\Gamma(\frac{5}{4}+\frac{p}{4})}{ \Gamma(\frac{7}{4}+\frac{p}{4}) } \cdot \frac{2^{\frac{p-1}{2}}}{p+1} \cdot \Gamma( \frac{p}{4}+\frac{19}{12} ) \Gamma( \frac{p}{4} - \frac{1}{12})\cdot x^{\frac{1-p}{2}} -\frac{\pi}{2} \frac{e^{-x}}{x}, \qquad x \gg 1.
\label{Q_large_equation}
\end{equation}

\begin{figure}
\begin{center}
\includegraphics[angle=-90, width=0.6\textwidth]{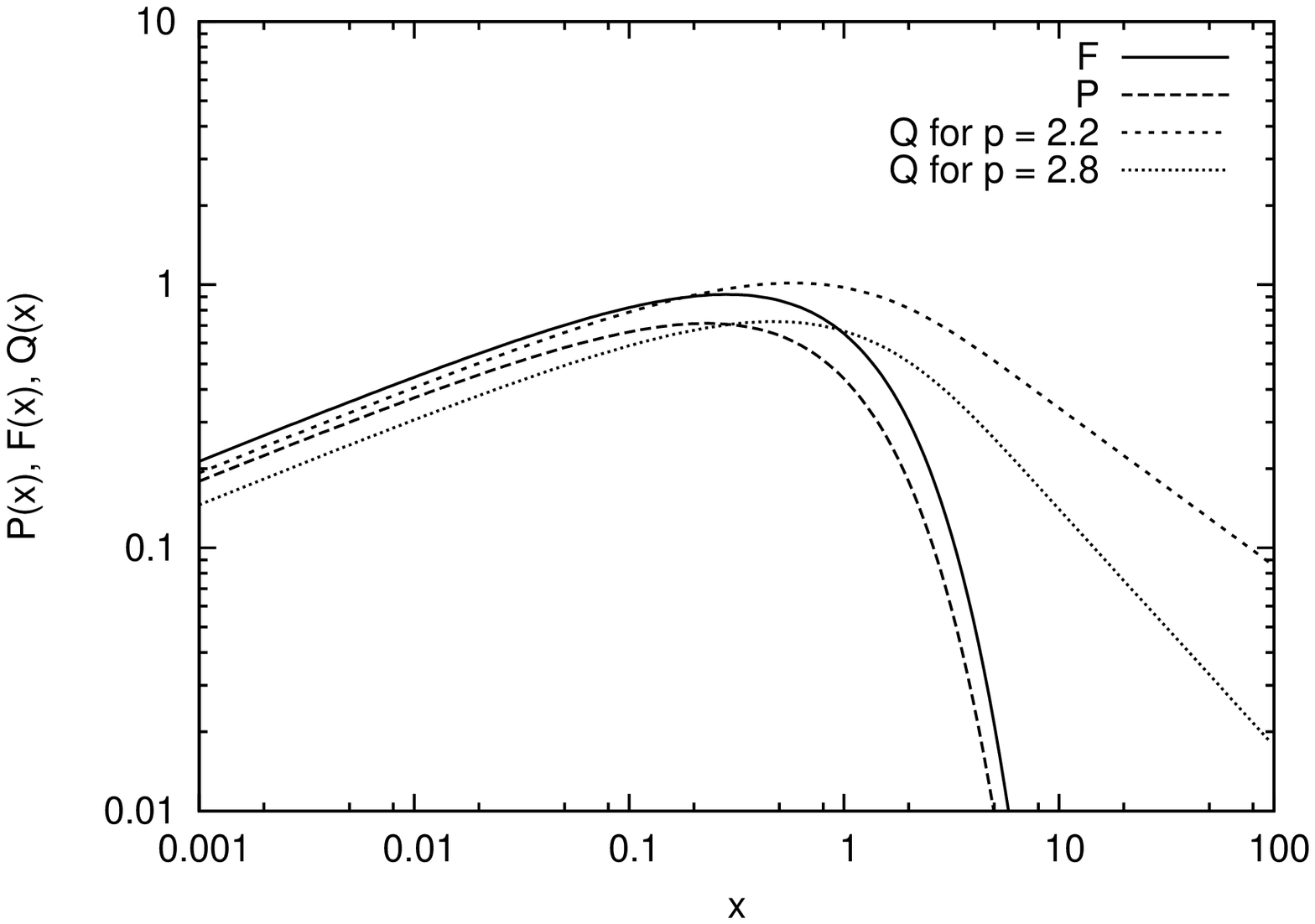}
\caption{$F(x)$, $\mathcal{P}(x)$ and $Q(x)$ (for $p = 2.2$ and $p = 2.8$ ): single electron, single angle-averaged electron and ensemble electron respectively.}
\label{FPQplot}
\end{center}
\end{figure}

In practice, the computer code uses lookup tables for $F(x)$, $\mathcal{P}(x)$ and $Q(x)$. The three functions have been plotted in figure (\ref{FPQplot}) ($Q$ for both $p = 2.2$ and $p = 2.8$), allowing for comparison between the spectra from a single electron, an angle-averaged electron and an ensemble electron.

\section{Emitted power with electron cooling}
\label{cooling_details_section}
If the only processes that are of importance are synchrotron emission and adiabatic cooling, the evolution of the Lorentz factor of a single electron is described by
\begin{equation}
\frac{\dev \gamma_e}{\dev t'} = - \frac{\sigma_T (B')^2}{6\pi m_e c} \gamma_e^2 + \frac{\gamma_e}{3n'} \frac{\dev n'}{\dev t'},
\label{evolution_equation}
\end{equation}
where $\sigma_T$ denotes the Thomson cross section. In \cite{Granot2002} this differential equation is applied to the BM solution by expressing it in terms of the self-similar variable and solving it analytically. In our case we can use eq. (\ref{gamma_m_equation}) to establish $\gamma'_m$ directly behind the shock front and initially put $\gamma'_M$, the upper cut-off Lorentz factor due to cooling, at a sufficiently large value (instead of infinity). Sufficiently large for example can be taken such that
\begin{equation}
\left| \frac{ \int_{\gamma'_m}^{\gamma'_M} \gamma_e^{1-p} - \int_{\gamma'_m}^{\infty} \gamma_e^{1-p} }{ \int_{\gamma'_m}^{\infty} \gamma_e^{1-p} } \right| \le \epsilon,
\end{equation}
with $\epsilon$ some tolerance for the error in the energy. The real $\gamma'_M$ will quickly catch up with the approximated $\gamma'_M$, as can be seen from equation (\ref{evolution_equation}).

The analytical solution for the particle distribution in the BM case is given by
\begin{equation}
 N_e (\gamma_e') = C \gamma_e'^{-p} \cdot ( 1 - \frac{\gamma_e'}{\gamma_M'})^{p-2},
\end{equation}
where the factor $C$ now stands for
\begin{equation}
 C = (p-1) \xi n' \gamma_m'^{p-1} \cdot ( 1 - \frac{\gamma_m'}{\gamma_M'} )^{1-p}.
\label{norm_equation}
\end{equation}
We take this to hold for the output of real RHD simulations as well, so that we have an approximate parametrisation for the particle distribution in any grid cell in terms of $\gamma'_m$ and $\gamma'_M$ alone. A more complete treatment of the particle distribution (e.g. \cite{Peer2005}) would effectively introduce an additional dimension to the RHD simulation and slow down the calculations accordingly.

Via reasoning completely analogous to the non-cooling case (where we use eq. (\ref{nume}) with $\gamma'_M$ instead of $\gamma'_e$ to obtain $\nu'_{cr,M}$) we arrive at an auxiliary function $\mathcal{Q}$ given by
\begin{equation}
 \mathcal{Q}(y_M, y_m) = y_m^{\frac{1-p}{2}}\cdot(1-(\frac{y_M}{y_m})^{1/2})^{1-p} \cdot \int_{y_M}^{y_m} y^{\frac{p-3}{2}} \cdot (1 - (\frac{y_M}{y} )^{1/2} )^{p-2} \mathcal{P}(y) \dev y,
\label{Q_cooling_equation}
\end{equation}
ocurring in
\begin{equation}
\frac{ \dev P'_{<e>}}{\dev \nu'} (\nu') = \frac{p-1}{2} \cdot \frac{\sqrt{3}{q_e}^3B'}{m_ec^2} \cdot 
    \mathcal{Q} \left( \frac{\nu'}{\nu'_{cr,M}}, \frac{\nu'}{\nu'_{cr,m}}\right).
\end{equation}
Since this is a function of two variables instead of one, its limiting behaviour is more complicated. If $y_M \ll 1$, $\mathcal{Q}(y_M, y_m )$ approximately reduces to
\begin{equation}
 \mathcal{Q}(y_M, y_m ) \propto (1 - ( \frac{y_M}{y_m} )^{1/2} )^{1-p} \cdot Q( y_m ), \qquad y_M \ll 1, 
\end{equation}
which can be obtained by approximating $y_M$ by zero in the integration limits and integrand of equation (\ref{Q_cooling_equation}). If $y_m / y_M \to 1$, the approximate result is
\begin{equation}
 \mathcal{Q}( y_M, y_m ) \propto \mathcal{P}(y_m), \qquad \frac{y_m}{y_M} \to 1,
\end{equation}
which follows from approximating the integral from (\ref{Q_cooling_equation}) by its value at $y_m$ times the integration domain. If $y_m \ll 1$ as well, we can use the first term of the lower limit series expansion for $\mathcal{P}$ in the integral and solve it to refine the approximate result to
\begin{equation}
 \mathcal{Q}( y_M, y_m ) \propto \mathcal{P} (y_m ) / (p - 1), \qquad \frac{y_m}{y_M} \to 1, \quad y_m \ll 1.
\label{Q_cool_equation}
\end{equation}

On the other hand, if $y_m / y_M \gg 1$, we can approximate the result in terms of $\mathcal{Q}( y_M, y'_m )$ for a smaller value $y'_m$ (i.e. the last tabulated value):
\begin{equation}
\mathcal{Q}( y_M, y_m ) \backsim \mathcal{Q}( y_M, y'_m ) \cdot \left( \frac{y_m}{y'_m} \right)^{\frac{1-p}{2}} + Q( y_m ) - Q( y'_m ) \left( \frac{y_m}{y'_m} \right)^{\frac{1-p}{2}}, \qquad \frac{y_m}{y_M} \gg 1,
\end{equation}
which further reduces to
\begin{equation}
\mathcal{Q}( y_M, y_m ) \backsim \mathcal{Q}( y_M, y'_m ) \cdot \left( \frac{y_m}{y'_m} \right)^{\frac{1-p}{2}} - \frac{\pi}{2} \frac{ e^{-y_m}}{y_m} + \frac{\pi}{2} \frac{e^{-y'_m}}{y'_m} \cdot \left( \frac{y_m}{y'_m} \right)^{\frac{1-p}{2}}, \qquad \frac{y_m}{y_M} \gg 1,
\end{equation}
for sufficiently high values of $y'_m$ and $y_m$.

Finally, for $y_M \gg 1$ we find from fitting to tabulated values that $\mathcal{Q}( y_M, y_m )$ is best described by
\begin{equation}
\mathcal{Q}( y_M, y_m ) \propto \left( \frac{ y_m }{y_M} \right)^{\frac{1-p}{2}} \cdot (1 - (\frac{y_M}{y_m})^{1/2} )^{(1-p)} \cdot e^{-y_M} \cdot y_M^{p-1}, \qquad y_M \gg 1.
\label{Q_drop_equation}
\end{equation}
In practice the code uses a two-dimensional table with numerically calculated values in addition to the analytical expressions above. The contribution from the region where $y_M \gg 1$ is effectively zero due to the exponential term $e^{-y_M}$.

\section{Derivation of Scaling coefficients}
\label{scaling_derivation_section}

We summarize the equations for the scaling coefficients in table (\ref{coefficients_table}). Aside from some explicit dependencies on $p$ and $k$ these equations also contain truly numerical constants with values that have been determined by fitting to output of our code. For example $C_D(p,k)$ contains the constants $C_{D0}$, $C_{Dk}$ and $C_{Dkk}$ (with $C_{Dk}^k$ denoting $C_{Dk}$ to the power $k$ etc.). Their numerical values are listed in table (\ref{coefficient_values_table}). Instead of incorporating these numerical constants in the total flux formula as we have done here we could also have used a fitting polynomial, but this approach more closely reflects the $k$ and $p$ dependencies.

\begin{table*}
 \centering
 \begin{minipage}{140mm}
\caption{scaling coefficients}
\begin{eqnarray*}
\hline
\hline
C_D & \equiv & (p-1) \cdot \left( C_{D0} C_{Dk}^k C_{Dkk}^{k^2} \right)^{1/(4-k)} \cdot \frac{1}{3-k} \cdot \left( \frac{p-2}{p-1} \right)^{-2/3} \\
 & & \cdot (17-4k)^{\frac{10-4k}{3(4-k)}} \cdot (4-k)^{\frac{2-k}{4-k}} \cdot Q_- \\
\hline
C_E & \equiv & (p-1) \cdot \left( C_{E0} C_{Ek}^k C_{Ekk}^{k^2} \right)^{1/(4-k)} \cdot \frac{1}{3-k} \\
  & & \cdot (17-4k)^{\frac{-6k+14}{3(4-k)}} \cdot (4-k)^{\frac{2-3k}{3(4-k)}} \cdot Q_{cool} \\
\hline
C_F & \equiv & (p-1) \cdot \left( C_{F0} C_{Fk}^k C_{Fkk}^{k^2} \right)^{1/(4-k)} \cdot \frac{1}{3-k} \\
 & & \cdot (17-4k)^{3/4} \cdot (4-k)^{-1/4} \cdot Q_{cool} \\
\hline
C_G & \equiv & (p-1) \cdot \left( C_{G0} C_{Gk}^k C_{Gkk}^{k^2} C_{Gp}^p C_{Gpk}^{pk} C_{Gpkk}^{pk^2} C_{Gpp}^{p^2} C_{Gppk}^{p^2k} C_{Gppkk}^{p^2k^2} \right)^{1/(4-k)} \cdot \frac{1}{3-k} \cdot \left( \frac{p-2}{p-1} \right)^{p-1} \\
 & & \cdot (17-4k)^{\frac{-kp-5k+4p+12}{4(4-k)}} \cdot (4-k)^{\frac{3kp-5k-12p+12}{4(4-k)}} \cdot Q_+ \\
\hline
C_H & \equiv & (p-1) \cdot \left( C_{H0} C_{Hk}^k C_{Hkk}^{k^2} C_{Hp}^p C_{Hpk}^{pk} C_{Hkk}^{pk^2} C_{Hpp}^{p^2} C_{Hppk}^{p^2k} C_{Hppkk}^{p^2k^2} \right)^{1/(4-k)} \cdot \frac{1}{3-k} \cdot \left( \frac{p-2}{p-1} \right)^{p-1} \cdot \\
 & & (17-4k)^{\frac{p+2}{4}} \cdot (4-k)^{\frac{2-3p}{4}} \cdot Q_+ \\
\hline
\hline
\end{eqnarray*}
\label{coefficients_table}
\end{minipage}
\end{table*}
The first term $(p-1)$ in these equations is also the first term in eq. (\ref{F_scaling_equation}). From the contribution of $N_{tot}$ we obtain a contribution $1/(3-k)$ via
\begin{equation}
N_{tot} = \xi_N 4 \pi \int_0^R r^2 n_0 \left( \frac{r}{R_0} \right)^{-k} \dev r
 = \xi_N \frac{4 \pi n_0 }{3-k} \left( \frac{R}{R_0} \right)^{3-k}.
\end{equation}
The origin of the combination $(p-2)/(p-1)$ can be traced to equation (\ref{gamma_m_equation}) in appendix \ref{distro_section} of this paper (\cite{Granot2002}) actually absorb it into $\epsilon_E$). The term $(17-4k)$ is linked to the energy $E_{52}$ and the two will always occur with the same power as can be seen from comparing tables \ref{coefficients_table} and \ref{scalings_table}. It enters our calculations via equation (69) from \cite{Blandford1976}. The term $(4-k)$ is likewise linked to the observer time $t_{obs,d}$. The extra term is a result from the transition from emission time in the grid lab frame to observer time. For the shock front the two are related via
\begin{equation}
t_{e} = ( 2 (4-k) t_{obs} )^{1/(4-k)} \left( \frac{E (17-4k)}{8\pi \rho_0 c^{5-k} R_0^k } \right)^{1/(4-k)}.
\label{t_e_front_equation}
\end{equation}
The final terms are different for the different power law regimes. They are contributed by the leading order terms of the various approximations of $\mathcal{Q}$. $Q_+$ is given by (see eqn. (\ref{Q_large_equation})):
\begin{equation}
Q_+ \equiv \frac{\Gamma(\frac{5}{4}+\frac{p}{4}) \Gamma ( \frac{p}{4}+\frac{19}{12} ) \Gamma ( \frac{p}{4}-\frac{1}{12} )}{ \Gamma(\frac{7}{4}+\frac{p}{4})(p+1)}.
\end{equation}
For low uncooled frequencies we have
\begin{equation}
Q_- \equiv \frac{1}{3p-1},
\end{equation}
as can be seen from equation (\ref{Q_small_equation}). When cooling plays a role we find that equation (\ref{Q_cool_equation}) provides us with
\begin{equation}
Q_{cool} \equiv \frac{1}{p-1}.
\end{equation}
Note that the effect of $Q_{cool}$ in $C_E$ and $C_F$ is to cancel out the first $(p-1)$ term -we only kept both terms for clarity of presentation.

\begin{table*}
 \centering
 \begin{minipage}{140mm}
\caption{Constants setting scale of flux}
  \begin{tabular}{|r|l|l|l|l|l|}
    \hline
       & D & G & H & F & E \\
    \hline
    0 & $5.12\cdot10^{-17}$ & $2.78 \cdot 10^{-31}$ & $5.68 \cdot 10^{-1}$ & $1.16 \cdot 10^{30}$ & $2.95 \cdot 10^{-16}$ \\
    $k$ & $1.18 \cdot 10^4$ & $4.54 \cdot 10^{7}$ & $6.94 \cdot 10^{-1}$ & $1.36 \cdot 10^{-8}$ & $2.04 \cdot 10^4$ \\
    $kk$ & $9.01 \cdot 10^{-1}$ & $8.95 \cdot 10^{-1}$ & $9.27 \cdot 10^{-1}$ & $1.01$ & $9.41 \cdot 10^{-1}$ \\
    $p$ & & $2.25 \cdot 10^{32}$ & $5.40 \cdot 10^{30}$ & & \\
    $pk$ & & $7.27 \cdot 10^{-9}$ & $1.65 \cdot 10^{-8}$ & & \\
    $pkk$ & & $9.41 \cdot 10^{-1}$ & $1.06$ & & \\
    $pp$ & & $1.77$ & $2.99$ & & \\
    $ppk$ & & $8.07 \cdot 10^{-1}$ & $7.01 \cdot 10^{-1}$ & & \\
    $ppkk$ & & $1.03$ & $1.01$ & & \\
   \hline
  \end{tabular}
\label{coefficient_values_table}
\end{minipage}
\end{table*}
\section{The hot region}
\label{hot_region_section}

For any given observer time and observer frequency there is a region behind the shock front where the emitting electrons have not yet cooled below the observer frequency. Although, when we set $\gamma_M$ initially at infinity, the size of this region never becomes zero, it can become very small, even when compared to the analytical error on the BM solution. The size of the hot region also determines the slope of the spectrum beyond the cooling break. We calculate its properties below.

The BM solution is obtained by a change of variables from $t$ and $r$ to $\chi$ and $1 / \Gamma^2$, where the fact that the latter becomes very small is continually used to simplify the dynamic equations using first order approximations. The $\chi$ coordinate of a fluid element is given by
\begin{equation}
\chi = [1 + 2(4-k)\Gamma^2] ( 1 - \frac{r}{ct} ),
\end{equation}
which is 1 at the shock front and increases roughly one order in magnitude until the back of the shock.

The radiation received at a given observer time is obtained by integrating over equidistant surfaces that have a one-on-one correspondence to emission times. To obtain an order of magnitude estimate for the size of the hot region we look solely at the jet axis, where each emission time and hence each equidistant surface corresponds to a given position $\chi$, via 
\begin{equation}
\chi( r, t ) = \chi( c (t_e - t_{obs}), t_e ) \approx \frac{ t_{obs}}{t_e} \cdot 2(4-k)\Gamma^2.
\label{chi_and_t_e_equation}  
\end{equation}
We define the boundary of the hot region $\chi_{hot}$ at the point where $\nu'_{cr,M} = \nu'$ (i.e. when the second argument of $\mathcal{Q}( \frac{\nu'}{\nu'_{cr,m}}, \frac{\nu'}{\nu'_{cr,M}} )$ is equal to one). The critical frequency $\nu'_{cr,M}$ is related to $\gamma'_M$ via the usual relation (see eqn. \ref{nume}), and an expression for $\gamma'_M$ in terms of the self-similar parameter $\chi$ can be found in \cite{Granot2002}:
\begin{equation}
 \gamma'_M(\chi) = \frac{2(19-2k)\pi m_e c \gamma}{\sigma_T B^2 t_e^2} \frac{1}{\chi^{(19-2k)/3(4-k)}-1}.
\end{equation}
Using the above we can find an expression for $\chi_{hot}$ -or equivalently $t_{e,hot}$, since the two are related via eqn. (\ref{chi_and_t_e_equation}). To first order in $\chi_{hot} - 1$ we find
\begin{equation}
 \chi_{hot} - 1 \approx \left( \frac{ 27 q_e m_e ( 4-k)^2}{ \nu \sigma_T^2 128 \sqrt{2 \pi} c^2 \epsilon_B^{3/2} \rho_0^{3/2} R_0^{3k/2} } \right)^{1/2} \cdot \left( \frac{E (17-4k)t_{obs} 2 (4-k)}{8 \pi \rho_0 R_0^k c^{5-k} } \right)^{\frac{4-3k}{4(k-4)}},
\end{equation}

\begin{equation}
 t_{e,front} - t_{e,hot} \approx \frac{1}{4-k} \cdot \left( \frac{ 27 q_e m_e ( 4-k)^2}{ \nu \sigma_T^2 128 \sqrt{2 \pi} c^2 \epsilon_B^{3/2} \rho_0^{3/2} R_0^{3k/2} } \right)^{1/2} \cdot \left( \frac{E (17-4k)t_{obs} 2 (4-k)}{8 \pi \rho_0 R_0^k c^{5-k} } \right)^{\frac{-3k}{4(k-4)}},
\end{equation}
where $t_{e,front}$ is the emission time of the shock front (see equation \ref{t_e_front_equation}).

From this we can draw a number of conclusions. The size of the hot region is dependent on the observer frequency via $\nu^{-1/2}$. For observer frequencies beyond the cooling break, it is effectively this region alone that contributes to the observed flux, since the contribution from the cool region drops exponentially (see equation \ref{Q_drop_equation}). A steepening of the spectral slope by -1/2 is therefore expected: $(1-p)/2 \to -p/2$. This results from multiplying the pre-cooling break flux by the fraction of the total emitting region that consists of the hot region -which is given by
\begin{equation}
\frac{t_{e,front} - t_{e,hot}}{t_{e,front}} \approx \frac{1}{4-k} \cdot \left( \frac{ 27 q_e m_e ( 4-k)^2}{ \nu \sigma_T^2 128 \sqrt{2 \pi} c^2 \epsilon_B^{3/2} \rho_0^{3/2} R_0^{3k/2} } \right)^{1/2} \cdot \left( \frac{E (17-4k)t_{obs} 2 (4-k)}{8 \pi \rho_0 R_0^k c^{5-k} } \right)^{\frac{4-3k}{4(k-4)}}
\end{equation}
Note that from this equation all post-cooling break scalings (e.g. $\epsilon_B$, $E_{52}$ etc.) can be derived by multiplying with the relevant pre-cooling flux.

Another important issue is that the size of the hot region can become smaller than the analytical error inherent in the BM solution, which cuts off beyond $1/\Gamma^2$. This happens at late times, when $\Gamma$ has dropped significantly. This puts a practical limit on a direct numerical implementation of the BM solution in our radiation code, ironically not due to numerical limitations of the code but because of the upper limit on the accuracy of the analytical solution that we have used to generate our grid files\footnote{In general, when post-processing grid files from simulations the issue does not occur because we use the AMR structure of the grid  to set the local integration accuracy. If the hot region becomes very small, then this will be dealt with at the earlier stage of the RHD simulation. Also, when directly integrating the flux equations for the BM solution by first expressing everything in terms of the self-similar coordinate and sticking to that frame, the issue is largely avoided as well.}. On can however still extrapolate the heuristic description of the spectra and light curves that we have obtained for arbitrary $k$ to late times -this is completely consistent with the canonical approach to light curve and spectrum modelling.

\bsp

\label{lastpage}

\end{document}